\documentclass[useAMS,usenatbib]{mn2e}
\usepackage{epsfig}

\def\co{{$^{56}{\rm Co}$}}

\def\ej{{\rm ej}}

\def\hh{{\rm H}}
\def\in{{\rm in}}
\def\iso{{\rm iso}}

\def\nk{{$^{56}{\rm Ni}$}}

\def\p{{\rm peak}}
\def\ph{{\rm ph}}
\def\rr{{\rm r}}

\def\x{{\rm X}}

\def\sw{{\it Swift}}

\title[XRF 080109 with a Normal Core-Collapse Supernova] 
{The X-Ray Transient 080109 in NGC 2770: an X-Ray~Flash Associated with a 
Normal Core-Collapse Supernova}
\author[Li-Xin Li]{Li-Xin Li\thanks{E-mail: lxl@mpa-garching.mpg.de}\\
Max-Planck-Institut f\"ur Astrophysik, 85741 Garching, Germany}

\begin{document}

\date{}


\date{Accepted 2008 May 13. Received 2008 April 28; in original form 2008 March 01}

\pagerange{\pageref{firstpage}--\pageref{lastpage}} \pubyear{2006}

\maketitle

\label{firstpage}

\begin{abstract}
Although it is generally thought that long-duration gamma-ray bursts 
(LGRBs) are associated with core-collapse supernovae (SNe), so far only 
four pairs of GRBs and SNe with firmly established connection have
been found. All the four GRB-SNe are among a special class of 
Type Ic---called the broad-lined SNe indicative of a large explosion 
energy, suggesting that only a small fraction of SNe Ibc have GRBs 
associated with them. This scheme has been refreshed by the discovery 
of a bright X-ray transient in NGC 2770 on 9 January 2008, which was 
followed by a rather normal Type Ib SN 2008D. In this paper, I argue 
that the transient 080109 is an X-ray flash (XRF, the soft version of a 
GRB) because of the following evidences: (1) The transient cannot be 
interpreted as a SN shock breakout event; (2) The GRB X-ray flare 
interpretation is not supported by the high-energy observation. Then I 
show that XRF 080109 satisfies the well-known relation between the 
isotropic-equivalent energy and the peak spectral energy for LGRBs, which 
highly strengthens the XRF interpretation. Finally, I point out that, the 
peak spectral energy of XRF 080109 and the maximum bolometric luminosity 
of SN 2008D agree with the $E_{\gamma,\p}$--$L_{{\rm SN},\max}$ 
relationship of Li (2006), strengthening the validity of the relationship.
I speculate that events like XRF 080109 may occur at a rate comparable to 
SNe Ibc, and a soft X-ray telescope devoted to surveying for nearby X-ray 
flares will be very fruitful in discovering them.
\end{abstract}


\section{Introduction}
\label{intro}

On 9 January 2008, a luminous X-ray transient was discovered during 
follow-up observations of SN 2007uy in NGC 2770 with the X-Ray Telescope (XRT)
onboard \sw\ \citep{ber08,kon08,sod08}. The transient event, which lasted 
about 600 s before decaying to the background level, was later confirmed to
occur in the same galaxy of SN 2007uy (at redshift $z =  0.006494$\footnote{
http://nedwww.ipac.caltech.edu/}) by the observation of an optical counterpart
and a supernova (SN), named SN 2008D, accompanying the transient (Deng \& Zhu
2008; Immler et al. 2008; Malesani et al. 2008a,b; Valenti et al. 2008a,b;
Blondin, Matheson \& Modjaz 2008; Thorstensen 2008; Li et al. 2008).

The SN underwent a transition from Type Ic to Type Ib \citep{mod08,val08b},
similar to the case of SN 2005bf \citep{wan05,mod05,anu05,tom05}. In addition,
the spectra of the SN exhibited rather broad features at early epochs, 
although not as broad as the earliest spectra of the so-called `hypernovae' 
associated with gamma-ray bursts (GRBs) such as SN 1998bw and SN 2006aj 
\citep{mal08,val08,blo08}. Despite these unusual characteristics compared
to typical SNe Ibc like 1994I, I call SN 2008D a `normal' SN as opposed to 
broad-lined GRB-SNe like 1998bw as suggested by other people 
\citep{mal08b,val08b,sod08}.

The nature of the transient 080109 is debatable. Several possibilities for
the nature have been proposed: (a) SN shock breakout \citep{bur08,sod08}; 
(b) X-ray flash (XRF, the soft version of a GRB) (Berger \& Soderberg 2008,
Xu, Zou \& Fan 2008); (c) X-ray flare of a GRB \citep{bur08}.

The total energy emitted by the transient is $\approx 1.3\times 10^{46}$ erg
in the XRT energy range $0.3$--$10$ keV (Section \ref{data}). Although this
energy is within the range predicted for shock breakout in SNe Ibc 
\citep{li07} and the transient occurred ahead of SN 2008D, the spectrum of the
transient is not a blackbody, and the event duration is too long for a typical
shock breakout event in a Type Ibc SN \citep{li07}. Hence, the shock breakout
interpretation does not seem to be correct (stronger arguments are given in 
Section \ref{shock}; see also Xu et al. 2008).

The transient 080109 was in the field of view of the Burst Alert Telescope 
(BAT) onboard \sw\ beginning half an hour before and continuing throughout 
the outburst. However, no gamma-ray counterpart was detected \citep{bur08}.
Hence, it is also unlikely that the transient was an X-ray flare of a GRB.

Then, the only remaining interpretation on the nature of the transient 080109
is that it is an XRF. In this paper I will show that, despite its extremely 
soft spectrum compared to that of a normal XRF or a GRB, the transient 080109
is naturally interpreted as an XRF in the context of the GRB-SN connection.

An important discovery in the observation of GRBs has been the connection 
between long-duration GRBs (with a duration $>2$~s) and SNe \citep[and 
references therein]{pir04,del06,woo06,nom08}. So far, four pairs of 
spectroscopically confirmed GRBs/SNe have been discovered: GRB 980425/
SN 1998bw \citep{gal98}, GRB 030329/SN 2003dh \citep{sta03,hjo03}, 
GRB 031203/SN 2003lw \citep{cob04,mal04,tho04}, and GRB 060218/SN 2006aj 
\citep{cam06,cob06,mir06,mod06,pia06,sol06,sod06}. All of the four SNe 
are among a special class of Type Ic, called the broad-lined SNe indicative
of a very large expansion velocity \citep[and references 
therein]{iwa98,maz03,del06,woo06}.

All the above four GRBs are nearby GRBs, among which GRB 030329 is the
farthest (at $z=0.17$). Observing SN signatures in 
high-redshift GRBs is difficult, since by selection effects the observable 
GRBs at high redshift are bright and hence the underlying SNe are easily 
overshone by the GRB afterglows. Despite this challenge, a handful of 
GRBs have shown rebrightening and flattening in their late optical afterglows,
which can be interpreted as the emergence of the underlying SN lightcurves
\citep{blo99,zeh04,sod05,bers06}. A systematic study on the GRB afterglows
with this approach suggests that all long-duration GRBs are associated with 
SNe \citep{zeh04}.

However, exceptions to the GRB-SN connection exist. Extensive observations of
two nearby long GRBs, 060614 at $z=0.125$ and 060505 at $z=0.089$, had not
detected SNe associated with them down to limits fainter than any SN Ic ever
observed \citep{del06b,fyn06,geh06}. This has been considered to be a 
challenge to the standard GRB classification scheme based on burst durations
\citep{zha06,wat07}.

On the other hand, whether a normal (not broad-lined) core-collapse SN
is associated with a GRB or a GRB-like event is uncertain. Considering the 
fact that GRBs are beamed so that many of them may have been missed by us, 
people have proposed to look for the GRB signature in nearby SNe by 
observing the late brightening in radio emissions of nearby SNe as 
expected when the GRB ejecta are slowed down and the radio emission becomes 
more or less isotropic \citep{pac01,lev02,gra03}. However, late-time radio 
observations of 68 local Type Ibc SNe, including six events with broad
optical absorption lines, have found none exhibiting radio emission 
attributable to off-axis GRB jets spreading into our line of sight 
\citep{sod06a}. This leads to a severe constraint on the fraction of SNe 
Ibc associated with normal GRBs.

With the four spectroscopically confirmed pairs of GRBs/SNe, a relation 
between the peak spectral energy of GRBs and the maximum bolometric 
luminosity or the mass of \nk\, in the ejecta of the underlying SNe 
was derived by \citet{li06}. A remarkable conclusion inferred from
the relation was that ``if normal Type Ibc SNe are accompanied by GRBs, the 
GRBs should be extremely underluminous in the gamma-ray band despite their 
close distances. Their peak spectral energy is expected to be in the soft 
X-ray and UV band, so they may be easier to detect with an X-ray or UV 
detector than with a gamma-ray detector.''(Li 2006, page 1362). For several
SNe Ibc that are not as luminous as SN 1998bw, the `expected GRB' derived 
from the relation has a peak spectral energy in the range of $0.01$--1 keV, 
and a total energy $10^{44}$--$10^{48}$ erg in the energy band 1-10000 keV.
It appears that XRF 080109 and SN 2008D agree with the relation (Section
\ref{xrf}).

The paper is arranged as follows. In Section \ref{data} the analysis of the 
XRT data is presented. In Section \ref{shock} it is argued that the X-ray 
transient 080109 cannot be interpreted as a SN shock breakout event. 
Section \ref{xrf} shows that the most natural explanation of the nature of 
the transient 080109 is that it is a faint XRF with a very soft spectrum. 
In Section \ref{model} models for producing faint and soft XRFs by a normal 
core-collapse SN are discussed. In Section \ref{concl} summary and 
conclusions are drawn, and future observational strategies are proposed.

\section{Data Reduction}
\label{data}

The XRT software was used to extract the lightcurve and the spectrum of
the X-ray transient 080109 in the XRT energy band $0.3$--10 keV from the 
Level 2 event data file (in Photon Counting mode) downloaded from the 
\sw\ online archive. 

The lightcurve has a FRED (Fast Rise and Exponential Decay) shape and a 
duration $\sim 600$ s. Although the X-ray emission was already in progress 
when the observation began and hence the start time of the burst is 
uncertain, from the shape of the lightcurve it is expected that the start 
time of the burst should not be too much earlier than the start time of 
observation (see, e.g., Fig. 1 of Soderberg et al. 2008).

In extraction of the source and background spectra over a time interval of
600~s containing the burst and beginning at the start time of observation, 
only events with grades 0--4 (i.e., single and double pixel events) were 
selected in order to achieve better spectral resolution. A fit of the King 
function to the point spread function (PSF) of the image showed that the 
core region with a radius of three pixels was piled up, so the core region 
was removed from the analysis. Then the source region where the spectrum was 
extracted was an annular aperture with an inner radius of four pixels 
($9.4^{\prime\prime}$) and an outer radius of 30 pixels ($71^{\prime\prime}$),
centered at the source position. The background region was defined as an 
annulus centered on the source with an inner radius of 60 pixels 
($142^{\prime\prime}$) and an outer radius of 110 pixels 
($260^{\prime\prime}$).  

In the data analysis, channels below 0.3 keV and above 10 keV were excluded.
The spectrum integrated over the 600~s time-interval was then fitted with the 
XSPEC package (v. 11.3.1), and the pile-up correction was included. To reduce 
the Poisson noises, the spectrum data were binned to ensure that each energy 
bin contains at least 20 counts.

The spectrum is well fitted by an absorbed power-law model, with an photon 
index $\Gamma = 2.29_{-0.26}^{+0.28}$ and total equivalent Hydrogen column 
density $N_\hh =6.83_{-1.3}^{+1.5}\times 10^{21}$ cm$^{-2}$. The reduced 
chi-square of the fit is $\chi_\rr^2 = 0.64$, with 18 degrees of freedom. 
These results are consistent with the analysis by \citet{sod08}. The fit
gives rise to an unabsorbed X-ray fluence of $1.36_{-0.37}^{+0.51}\times 
10^{-7}$ erg cm$^{-2}$ in $0.3$--10 keV. [A figure of the power-law
spectral fit is not shown here since it is very similar to the fig. 2
of Soderberg et al. (2008).]

The fitted Hydrogen column density is much higher than the Galactic Hydrogen 
column density to the direction of NGC 2770 \citep[$2\times 10^{20}$ 
cm$^{-2}$,][]{kon08}, but is typical among the Hydrogen column density of 
GRB host galaxies \citep{sav06}.

Given the redshift of the source, $z=0.0065$, the power-law spectral fit
leads to an unabsorbed total energy $1.27^{+0.48}_{-0.35}\times 10^{46}$ erg
and an unabsorbed average luminosity $2.11^{+0.79}_{-0.58}\times 10^{43}$ erg
s$^{-1}$ in the XRT's energy range $0.3$--$10$ keV. The luminosity
is $10^5$ times of the Eddington luminosity of a solar mass object.

\begin{figure}
\vspace{2pt}
\includegraphics[angle=-90,scale=0.344]{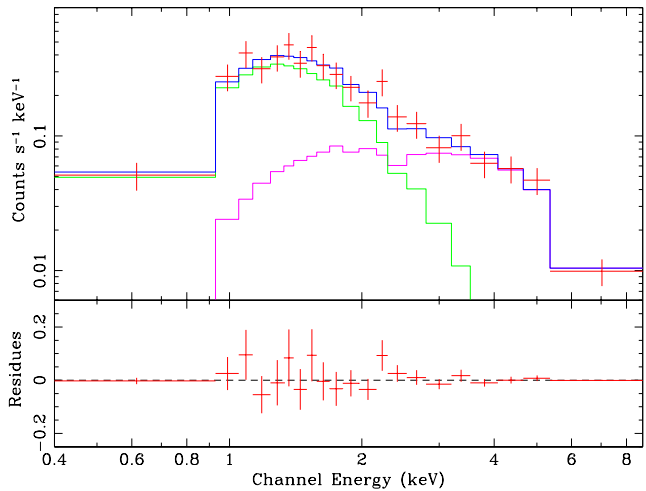}
\caption{Fit of the X-ray spectrum ($0.3$--10 keV) of XRF 080109 by an 
absorbed double blackbody model (blue line). The best fit is given by 
$N_\hh = 4.7\times 10^{21}$ cm$^{-2}$, $T_1 = 0.36$ keV (green line), and 
$T_2 = 1.24$ keV (magenta line). The reduced chi-square of the fit is 
$\chi_\rr^2 = 0.65$, with $16$ degrees of freedom.
}
\label{xrf080109_bb_bb}
\end{figure}

Although the power-law fit is good, the fit of the spectrum is not unique. 
An interesting result is that, although an absorbed single blackbody does 
not fit the spectrum of XRF 080109 ($\chi_\rr^2 = 2.2$), an absorbed double 
blackbody model fits the spectrum equally well as the absorbed power-law 
(Fig.~\ref{xrf080109_bb_bb}). The fitted parameters are: $N_\hh = 
4.7_{-1.8}^{+2.6} \times 10^{21}$ cm$^{-2}$, the temperature of the first 
blackbody component $T_1 = 0.36_{-0.094}^{+0.12}$ keV, and the temperature 
of the second blackbody component $T_2 = 1.24_{-0.24}^{+0.52}$ keV. The 
reduced chi-squares of the fit is $\chi_\rr^2 = 0.65$, with $16$ degrees 
of freedom. 

With this double blackbody model, the unabsorbed total bolometric energy
radiated by the burst is $7.31_{-1.3}^{+2.4} \times 10^{45}$ erg, with 
$3.49_{-1.03}^{+2.31} \times 10^{45}$ erg in the lower temperature component, 
and $3.82_{-0.78}^{+0.80} \times 10^{45}$ erg in the higher temperature 
component.

I stress that the above result only indicates that a blackbody origin of the
X-ray emission cannot be ruled out, and it does not prove that a blackbody
component exists in the X-ray emission.

\section{Is the Transient 080109 a Supernova Shock Breakout Event?}
\label{shock}

The X-ray spectrum of the 080109 transient can be fitted with a power law 
(Section \ref{data}). No blackbody component is required. A characteristic 
feature of SN shock breakout is a blackbody-like spectrum 
\citep{ims89,mat99}. Hence, the transient event 080109 is not likely to be
a SN shock breakout event.

However, as shown in Section \ref{data}, the spectrum can be equally well 
fitted with a model consisting of two blackbody components 
(Fig.~\ref{xrf080109_bb_bb}). Now let us check if one of the two blackbody 
components might arise from the SN shock breakout event.

From the duration of the event and the bolometric total energy of each
blackbody, the average bolometric luminosity and hence the average 
photosphere radius of each component can be derived. For the softer component
($T_1 = 0.36$ keV), the radius is $R_\ph \approx 0.074 R_\odot$. For the 
harder component ($T_2 = 1.24$ keV), the radius is $R_\ph \approx 0.0062
R_\odot$. Both are much smaller than the solar radius. 

The underlying SN of the event, SN 2008D, was initially classified as 
Type Ic and later reclassified as Type Ib \citep{mod08,val08b}. This indicates
that the progenitor star should be a Wolf-Rayet star, which usually has a 
radius of several solar radii. SN shock breakout occurs at a radius near the 
stellar surface \citep{ims89,mat99,tan01}, or the photospheric radius if the 
star is surrounded by an intense stellar wind \citep{li07}. Hence, the above 
results on the photospheric radius of the blackbody emission indicate that 
neither of the two blackbody components originates from the shock breakout 
event.

From the derived photospheric radii, a limit on the expansion speed of the
blackbody photosphere can be estimated. Assume that the average photospheric
radius corresponds to a time of 100~s after the explosion. Then, for the 
softer blackbody component, the photospheric speed $v_\ph \la 0.0017 c$
(where $c$ is the speed of light). 
For the harder blackbody component, the photospheric speed $v_\ph \la 
0.00015 c$. These speeds are non-relativistic, contrary to the 
prediction that shock breakout from a compact Wolf-Rayet star is mildly 
relativistic \citep[with a shock breakout velocity $\ga 0.3 c$,][]{tan01,li07}.

The duration of the X-ray transient 080109 is $\approx 600$~s, which is also
much longer than that predicted for shock breakout in SNe Ibc. Hence, we 
conclude that the X-ray transient 080109 is not a SN shock breakout 
event.

\section{Transient 080109 as an X-Ray Flash from a Normal Core-Collapse
Supernova}
\label{xrf}

Having shown that the X-ray transient 080109 is not a SN shock 
breakout event, one is left with two alternative interpretations about the
nature of the transient: (1) It is a low-luminosity XRF \citep{ber08,xu08};
(2) It is a flare in the X-ray afterglow of a GRB \citep{bur08}. 

The transient object happened to be in the BAT field of view in two previous
\sw\ observations (of BZQ J0618+4620 beginning at 13:04:12.33, and of 
SN 2007ax beginning at 13:12:24.5 UT on 9 Jan 2008). BAT did not trigger 
during either of the two observations. An examination of the BAT data from 
the direction of NGC 2770 during those observations shows no sign of emission
in the BAT energy range 15--150 keV, with a fluence upper limit of $\sim 
1.0\times 10^{-7}$ erg cm$^{-2}$ in a period of half an hour before the start
of observation of the transient 080109 \citep{bur08}. In addition, the UVOT
lightcurve of the transient closely resembles an early stage UV-optical
lightcurve of a GRB (e.g., GRB 060218), rather than a UV-optical afterglow 
lightcurve during the late declining stage. Hence, the interpretation of the 
transient event as a flare in the X-ray afterglow of a GRB is also ruled out.

An additional evidence supporting an XRF interpretation of the transient
080109 is in the shape of the X-ray lightcurve after the prompt emission
phase. At the end of an exponential decay, the X-ray lightcurve breaks to a
power-law decay with an index $\approx -1.1$ up to $t\approx 30000$~s from
the start of observation, which is a characteristic of typical GRB afterglows
\citep{xu08}.

Based on the above arguments, I conclude that the transient event 080109 in
NGC 2770 is a soft XRF, and XRF 080109/SN 2008D well fits the framework of 
the GRB-SN connection.

\begin{figure}
\vspace{2pt}
\includegraphics[angle=0,scale=0.468]{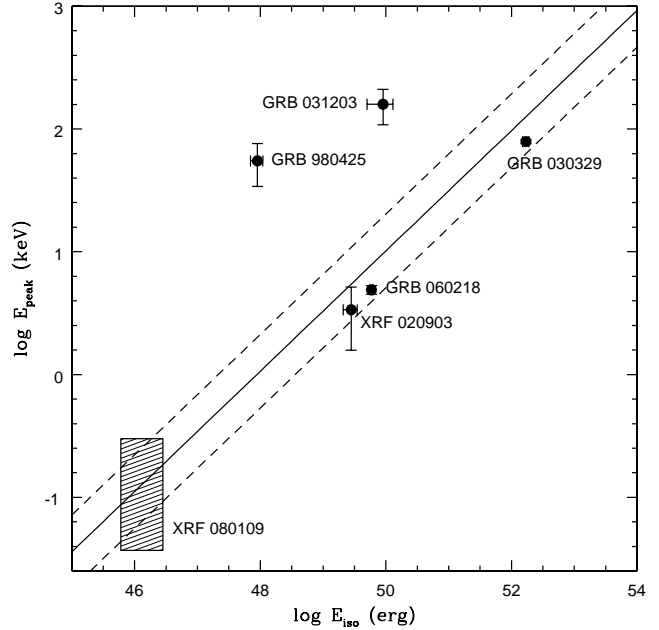}
\caption{The peak spectral energy versus the isotropic equivalent energy
in 1--10000 keV for nearby long-duration GRBs. The solid line is the
$E_\iso$--$E_\p$ relation obtained from a power-law fit to 41 long-duration
cosmological GRBs \citep{ama06}. The two dashed lines delineate the region 
of a logarithmic deviation of $0.3$ (2-$\sigma$) in $E_\p$. For XRF 080109,
$E_\iso\approx 1.3_{-0.7}^{+1.5}\times 10^{46}$ erg, and $E_\p$ is in the 
range of $0.037$--$0.3$ keV (see text). XRF 080109 appears to be consistent
with the Amati relation. GRBs 980425 and 031203 are well-known outliers to
the $E_\iso$--$E_\p$ relation.
}
\label{eiso_epeak2}
\end{figure}

XRF 080109 is more under-luminous than the previous most under-luminous
burst, GRB 980425, by about two orders of magnitude. The spectrum of 
XRF 080109 is also softer than that of GRB 980425. During the XRT observation
of XRF 080109 which lasted over 1000~s, the BAT fluence upper limit is $8.9
\times 10^{-8}$ erg cm$^{-2}$ in 15--150 keV \citep{bur08}. Extrapolation of 
the power-law spectral fit in Section \ref{data} to 15--150 keV leads to a 
fluence of $3.4_{-2.2}^{+6.3}\times 10^{-8}$ erg cm$^{-2}$, marginally 
consistent with the BAT upper limit.

Due to the limit in the number of photon counts (433 in total after the 
piled-up core region being removed) and the small range of energy covered by
XRT ($0.3$--10 keV), a reliable constraint on the peak spectral energy cannot
be obtained from the XRT data alone. However, the fact that the XRT spectrum 
of XRF 080109 can be fitted by a single power-law with a 
photon index $\Gamma\approx 2.3$ suggests that the peak spectral energy 
$E_\p<0.3$ keV. A lower limit on the value of $E_\p$ can be obtained from 
the UVOT observation during the prompt phase of XRF 080109. The specific flux
density in the {\em UBV} band (at $\sim 3$~eV) during the prompt phase is 
$F_\nu<9.0\times 10^2$ $\mu$Jy \citep{imm08,sod08}. According to the 
synchrotron model for GRB emissions \citep{sar98}, $F_\nu\propto \nu^\alpha$
with $\alpha\le 1/3$. Then, combination with the power-law fit to the XRT 
spectrum leads to a constraint on $E_\p$: $E_\p > 0.037$ keV. Hence we have 
$0.037~{\rm keV} < E_\p <0.3~{\rm keV}$.

The power-law spectral fit leads to a total isotropic-equivalent energy
$E_\iso = 1.3_{-0.7}^{+1.5}\times 10^{46}$ erg in 1--10000 keV in the rest
frame of the burst. This value of $E_\iso$, together with the constraint on
the peak spectral energy obtained above, makes XRF 080109 agree with the 
$E_\iso$--$E_\p$ relation of \citet{ama06} (see Fig. \ref{eiso_epeak2}). 
Alternatively, from the value of $E_\iso$ for XRF 080109, the Amati relation 
implies that the peak spectral energy of XRF 080109 should be $E_\p \approx 
0.12_{-0.089}^{+0.23}$ keV, in good agreement with the constraint inferred
from the XRT and UVOT data.

\begin{figure}
\includegraphics[angle=0,scale=0.474]{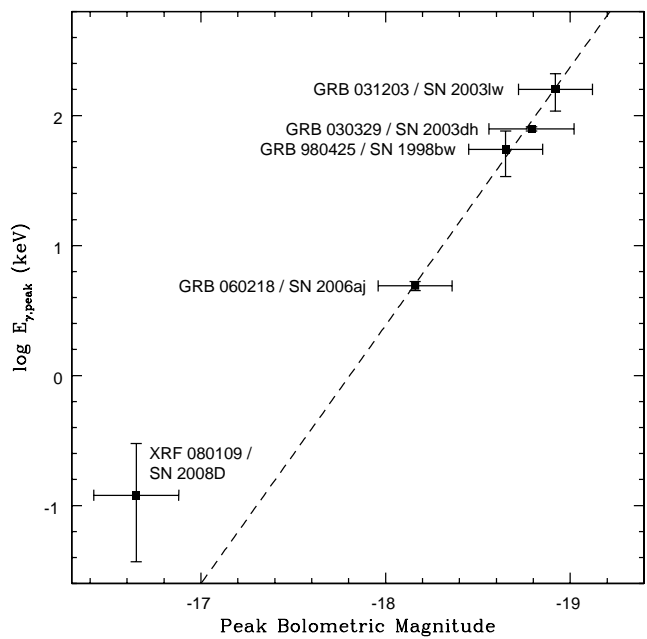}
\caption{Relation between the peak spectral energy of GRBs and the peak
bolometric magnitude of the underlying SNe (Li 2006, with the addition of
XRF 080109/SN 2008D). The straight dashed line is the best fit to the four 
pairs of GRBs/SNe (980425/1998bw, 030329/2003dh, 031203/2003lw, and
060218/2006aj) with spectroscopically confirmed connection (eq. 1 in Li 
2006).
}
\label{mag_epeak}
\end{figure}

\begin{figure}
\includegraphics[angle=0,scale=0.474]{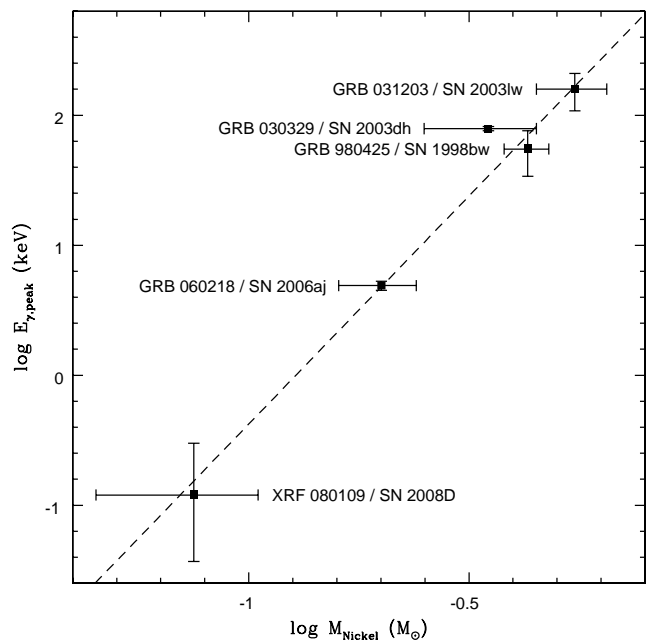}
\caption{Relation between the peak spectral energy of GRBs and the mass of
\nk\, generated in the ejecta of the underlying SNe (Li 2006, with the
addition of XRF 080109/SN 2008D). The straight dashed line is the best fit 
to the four pairs of GRBs/SNe with spectroscopically confirmed connection 
(caption of fig. 3 in Li 2006).
}
\label{n56_epeak}
\end{figure}

The bolometric lightcurve of SN 2008D in the early stage was derived from
the UVOT data and modeled by \citet{sod08}. The peak of the lightcurve 
occurred at about 20~day after the explosion, with a peak bolometric magnitude 
$\approx -16.65$ (corresponding to a maximum bolometric luminosity $\approx 
1.4\times 10^{42}$ erg s$^{-1}$). Fitting the lightcurve by an analytic model
of SN emission powered by the radioactive decay of \nk\, and \co\,
yielded a \nk\ mass synthesized in the explosion between $0.05$ and 
$0.1 M_\odot$ \citep{sod08}. These results, 
together with the peak spectral energy of XRF 080109 derived from the XRT
and UVOT data, indicate that XRF 080109/SN 2008D agree with the relation 
between the peak spectral energy of GRBs and the maximum bolometric 
luminosity or the nickel mass in the ejecta of the underlying SNe 
derived by \citet{li06}, as shown in Figs. \ref{mag_epeak} and 
\ref{n56_epeak} (where the central value of $E_\p$ is 0.12~keV estimated
by the Amati relation).

\begin{figure}
\vspace{2pt}
\includegraphics[angle=0,scale=0.475]{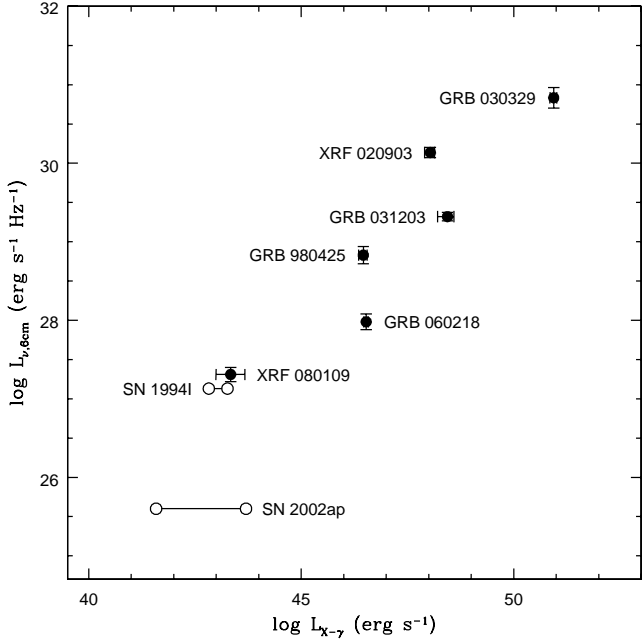}
\caption{The peak radio luminosity at 6~cm versus the average luminosity
in the X-ray/gamma-ray band for several local GRBs/XRFs and normal Type Ibc
SNe. SN 1994I and SN 2002ap do not have detected GRBs/XRFs. The 
X-ray/gamma-ray luminosity of the potential bursts associated with them was 
derived from their peak bolometric luminosity (left circle) and their nickel
mass (right circle; see text for details). References for the radio 
luminosity are: 
\citet{bjo04} (SN 2002ap), \citet{hor07} (GRB 030329), \citet{sod04a} 
(XRF 020903), \citet{sod04b} (GRB 031203), \citet{sod06} (XRF 060218), 
\citet{sod08} (XRF 080109), and \citet{wei02} (GRB 980425, SN 1994I).
}
\label{grb_radio}
\end{figure}

All known nearby GRBs/XRFs with SNe have strong radio emissions, 
including GRBs 980425, 030329, 031203, 060218, and XRF 020903.\footnote{A
supernova has been claimed being detected in the afterglow of XRF 020903
\citep{sod05,bers06}, which is fainter than SN 1998bw by 0.6--0.8 mag at 
the peak in the $R$-band. However, the bolometric magnitude could not be
obtained due to the lack of data in other filters.} Radio emissions have 
also been detected for XRF 080109/SN 2008D, although not as bright as 
the other GRB-SNe \citep{sod08}. The peak radio luminosities at 6~cm
($L_{\nu,6{\rm cm}}$) of the six GRBs/XRFs are plotted in 
Fig.~\ref{grb_radio}, versus their average luminosity of the prompt 
emission in the X-ray and gamma-ray band ($L_{\x-\gamma}$). The data suggest
a correlation between the radio luminosity and the X-ray/gamma-ray 
luminosity, i.e., brighter GRBs/XRFs tend to have a larger radio 
luminosity.

Type Ic SN 1994I and SN 2002ap were also detected in radio band, although no 
GRBs/XRFs have been found to be associated with them. With the relation
between the peak spectral energy of GRBs and the maximum bolometric 
luminosity of the underlying SNe (or the mass of \nk\, generated in 
the SN ejecta), the peak spectral energy of the potential XRFs 
associated with SN 1994I and SN 2002ap was derived to be $0.07$ keV (or 
$0.12$ keV) and $0.016$ keV (or $0.19$ keV), respectively \citep{li06}. 
Using the Amati relation, the peak spectral energy can be converted to the
isotropic-equivalent energy in the 1-10000 keV band. Assuming a duration 
of 600~s (the same duration of XRF 080109) for these potential faint bursts, 
the average luminosity of the prompt emission in the X-ray/gamma-ray band can 
be calculated. The peak radio luminosities of SN 1994I and SN 2002ap and 
the derived average luminosities of their potential bursts in the 
X-ray/gamma-ray band are shown Fig. \ref{grb_radio} by open circles. It 
appears that they follow the trend of the $L_{\nu,6{\rm cm}}$--$L_{\x-\gamma}$
relation suggested by the nearby GRBs/XRFs.

\section{How Are Spherical GRBs/XRFs Produced?}
\label{model}

In the standard collapsar model of long-duration GRBs (MacFadyen \& Woosley 
1999; MacFadyen, Woosley \& Heger 2001), it is assumed that after the 
core-collapse 
of the progenitor star a torus is formed surrounding a rapidly rotating black
hole. A bipolar relativistic fireball outflow powered either by the accretion 
energy of the torus or the spin energy of the black hole is generated, and
collimated into two oppositely-directed jets moving along the spin axis of the
hole. The fireball is presumably highly nonhomogeneous and composed of a 
number of outward moving shells. The collision between the shells produces 
the prompt gamma-ray emission, and the collision between the shells and the 
surrounding medium produces the afterglow emission
\citep[the so-called internal/external-shock model,][]{pir04}. In this model, 
collimation of the outflow is essential for avoiding baryon loading and
maintaining a large Lorentz factor ($> 100$).

However, it appears that some GRBs are spherical. An investigation on the
relation between the jet opening angle and the peak spectral energy of GRBs
revealed that they are anti-correlated \citep{lam05,li06}. That is, GRBs with
softer spectra have larger jet opening angles i.e. weakly collimated outflows.
For GRBs/XRFs with a peak spectral energy $<40$ keV (in the burst frame),
the jet opening angle inferred from the anti-correlation is so large that
the burst outflow should be spherical \citep{li06}. This is consistent with 
the radio observation on XRF 020903 ($E_\p \approx 3.4$ keV) and GRB 060218
($E_\p \approx 4.9$ keV) \citep{sod04a,sod06}. XRF 080109 has a spectrum
softer than that of XRF 020903 and GRB 060218, hence the outflow of it should 
also be spherical. This is consistent with the conclusion that SN 2008D has
a non-relativistic ejecta as inferred from its radio emission \citep{sod08}.

Obviously, the standard internal/external-shock fireball model does not apply 
to GRBs/XRFs with a spherical outflow, since a spherical fireball outflow 
cannot avoid baryon loading efficiently: it must pass through the dense
SN ejecta. Then, an unavoidable result of a spherical GRB/XRF is that
the outflow which produces the burst and the afterglow cannot have a very 
large Lorentz factor. Due to the loss of energy to the SN ejecta, the
burst would be very sub-energetic compared to normal GRBs. In other words, 
a spherical GRB/XRF would have a low luminosity and soft spectrum, and be 
at most mildly relativistic.

Whether a spherical explosion can produce a GRB-like event is a question.
The GRB fireball is trapped inside the heavy SN envelope so the energy
of it may well be dissipated by the SN envelope without producing
a GRB/XRF. However, two possible scenarios for producing a GRB/XRF from a
spherical configuration can be imagined.

{\em Scenario A.} When a light fluid is accelerated into a heavy fluid, which
is just the case of a spherical GRB explosion as outlined above, the 
Rayleigh-Taylor instability occurs \citep{dra04}. The Rayleigh-Taylor 
instability has been proposed as a mechanism for driving SN explosion
\citep{buc80,sma81}, and a mechanism for the mixing of elements in SN 
explosion \citep{hac90,hac91,mue97}. In the case of a spherical GRB/SN
explosion, the GRB fireball may emerge from the SN envelope through 
the Rayleigh-Taylor instability, then produce a GRB/XRF through either 
the internal-shock or the external-shock interaction.

{\em Scenario B.} The initial GRB fireball is killed by the SN 
envelope and the fireball energy is added to the SN explosion energy. 
A small fraction of the outer layer of the SN envelope is accelerated 
by the enhanced SN shock wave to a mildly relativistic velocity and 
generates a low-luminosity GRB/XRF via interaction with surrounding matter. 
This GRB-production mechanism through acceleration of the SN outer
layer has been proposed by \citet{mat99} and \citet{tan01} for explaining 
the prompt emission of GRB 980425. Applying the formulae for this 
mechanism \citep{tan01} to SN 2008D, with an assumption of the SN 
explosion energy $E_\in \approx 3\times 10^{51}$ erg, the ejected mass $M_\ej
\approx 4 M_\odot$ \citep{sod08}, and the progenitor mass $M_\star \approx 
6 M_\odot$, a total kinetic energy $\approx 4\times 10^{46}$ erg is obtained 
for ejecta with a velocity $>0.5 c$. This energy is enough to account for the 
total X-ray energy emitted by XRF 080109. However, this mechanism is not able
to explain GRB 060218, which has a total isotropic energy $\approx 6\times
10^{49}$ erg emitted in X-ray/gamma-ray, exceeding the predicted kinetic 
energy by three orders of magnitude [with $E_\in \approx 2\times 10^{51}$
erg, $M_\ej\approx 2 M_\odot$, and $M_\star\approx 3.3 M_\odot$ adopted from 
\citet{maz06}].

\section{Summary and Conclusions}
\label{concl}

\begin{figure}
\vspace{2pt}
\includegraphics[angle=0,scale=0.463]{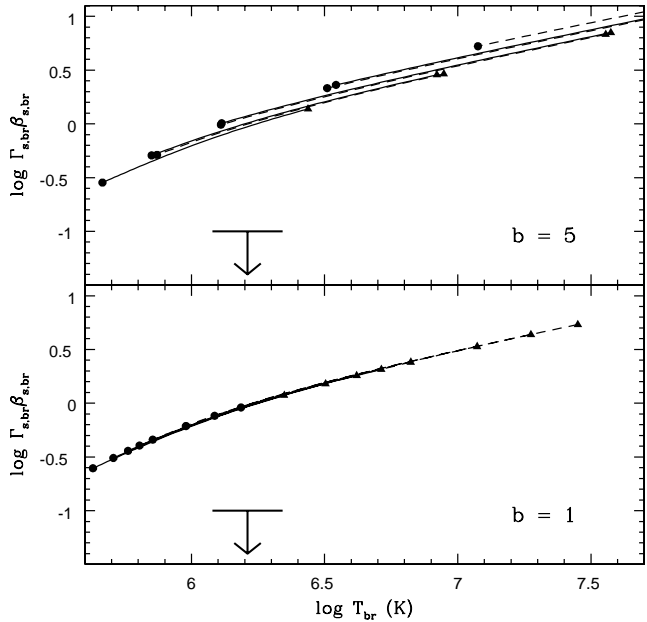}
\caption{The predicted momentum of shock wave at breakout for SN 2006aj,
versus the breakout temperature (for the meaning of parameters please refer
to Li 2007a). Solid lines correspond to opacity $\kappa_w=0.2$ cm$^2$ 
g$^{-1}$. Dashed lines correspond to $\kappa_w=1$ cm$^2$ g$^{-1}$. Different 
solid lines (and different dashed lines) in the same panel correspond to 
different values of parameter $\varepsilon$: $10^{-1}$, $10^{-2}$, $10^{-3}$ 
and $10^{-4}$ (upward). Along each line the stellar radius $R_\star$ varies 
from $1 R_\odot$ (the triangle) to $100 R_\odot$ (the point). It is assumed 
that the SN involves explosion energy $E_\in = 2\times 10^{51}$ erg 
and ejected mass $M_\ej = 2 M_\odot$ \citep{maz06}. The horizontal line with
a downward arrow denotes the measured temperature and photosphere velocity 
of GRB 060218.
}
\label{tembr_gsvs}
\end{figure}

The X-ray transient event 080109 in NGC 2770 is the first-ever X-ray flash 
discovered in a normal core-collapse supernova. It emitted an energy of $E_\x
\approx 1.3 \times 10^{46}$ erg in the \sw\ XRT's $0.3$--10 keV band in a 
duration of $\sim 600$ s. The XRT spectrum of XRF 080109 is well fitted by 
an absorbed power-law model with a photon index $\approx 2.3$. In combination 
with the upper limit of the UVOT observation during the prompt emission phase, 
the XRT spectral fit leads to a constraint on the peak spectral energy of 
XRF 080109: $0.037~{\rm keV}<E_\p<0.3$~keV. The total isotropic-equivalent 
energy in 1--10000 keV in the rest frame of the burst is $E_\iso\approx 1.3
\times 10^{46}$ erg. With the above values of $E_\iso$ and $E_\p$, XRF 080109
is consistent with the Amati $E_\iso$--$E_\p$ relation.

Although the XRT spectrum of XRF 080109 can also be fitted with an absorbed
double blackbody model, the photospheric radius derived from the luminosity 
and the temperature of each blackbody component rules out a SN shock 
breakout interpretation for the burst. For SN shock breakout 
from a Wolf-Rayet progenitor star, the radius where the shock breakout occurs
is expected to be not smaller than several solar radii. But the photospheric 
radii derived for the two blackbody components are much smaller than the 
solar radius. Similarly, for the case of GRB 060218, the shock breakout
interpretation for the blackbody component in the early X-ray and UV-optical
emissions \citep{cam06} is not valid because of the extremely long duration
and large energy of the blackbody component \citep{li07} as well as the
inconsistence of the calculated photospheric velocity with the model 
(Fig.~\ref{tembr_gsvs}).

In combination with the fact that the high energy observation by the BAT/\sw\
has ruled out a GRB X-ray flare interpretation, the above facts strongly 
support the claim that the X-ray transient 080109 is an XRF from a normal 
core-collapse supernova.

The SN associated with it, SN 2008D, is a normal Type Ibc SN in terms of its 
peak luminosity and the optical spectra, in contrast to other GRB-SNe like 
SN 1998bw. Adopting the peak bolometric luminosity and the mass of nickel 
in the ejecta of SN 2008D derived by \citet{sod08}, XRF 080109/SN 2008D is 
consistent with the relation between the peak spectral energy of GRBs and the
peak bolometric luminosity or the mass of nickel of the underlying SNe 
proposed by \citet{li06}. The radio observation on XRF 080109 puts it in a 
place in the plane of the X-ray/gamma-ray luminosity versus the radio 
luminosity that agrees with the overall trend of nearby GRBs/SNe.
 
The very soft spectrum of XRF 080109 indicates that the fireball outflow
associated with it is more or less spherical. How a spherical fireball plows
through a dense environment of the SN envelope and makes a `spherical
GRB' is discussed. In addition to the model that a fraction of mass in the 
outer layer of the SN envelope is accelerated by the shock
wave to a mildly relativistic speed which then produces a faint GRB via
interaction with the surrounding matter \citep{tan01}, a new scenario is 
proposed which assumes that the low-density fireball emerges from the 
SN envelope via the Rayleigh-Taylor instability. During its
way of pushing the heavy envelope aside, the fireball loses energy and gets 
mass from the SN envelope. Then an intrinsically faint and 
spectroscopically soft burst is produced by the mildly relativistic or
sub-relativistic fireball via the internal/external shock interaction.

The detection of an XRF in a normal core-collapse SN extends the connection 
between GRBs/XRFs and SNe. It may suggest that every Type Ibc (maybe Type II
also) SN has a GRB/XRF associated with it. If this is true, events like 
XRF 080109 would occur at a rate comparable to that of SNe Ibc, $\sim 
10^{-3}$ yr$^{-1}$ in an average galaxy \citep{pod04}.

For a normal core-collapse SN that is not as bright as SN 1998bw, it
is expected that the XRF associated with it has a spectrum peaked in the UV 
to X-ray band and total energy of $10^{44}$--$10^{48}$ erg \citep{li06}. 
A wide-field soft X-ray telescope devoted to the detection of soft X-ray 
flares will be very fruitful in discovering these very soft XRFs. The 
detection of them is very important for understanding the nature of the GRB-SN 
connection, the emission mechanism of low-luminosity XRFs, and the explosion
mechanism of core-collapse SNe.

The design and launch of such a telescope is also important for detecting 
other three types of events: (1) Shock breakout in SNe Ibc \citep{li07},
which is the first observable electromagnetic signal from a core-collapse
SN and should occur ahead of the XRF; (2) Thermal precursors of 
normal GRBs \citep{li07b}, the detection of which will give us a chance
to obtain a complete multi-wavelength observation on GRBs starting from
the prompt emission phase \citep[e.g.,][]{cen06}; (3) Bright GRBs at very
high redshift ($z>10$) whose peak spectral energy is shifted to the soft
X-ray range by the cosmic redshift effect.

\section*{Acknowledgments}

The author thanks Bin-Bin Zhang for help in data reduction, Jinsong Deng, 
Yi-Zhong Fan, Daniele Malesani, and Bing Zhang for useful communications.
He also thanks the referee (Elena Pian) for a very helpful report which has
led to significant improvements to the presentation of the paper.

\bsp

\label{lastpage}

\end{document}